\documentclass[12pt]{article}
\usepackage{graphicx}
\usepackage{amsfonts}
\usepackage{amssymb,amsmath}
\usepackage{color}
\usepackage[colorlinks=true,linkcolor=blue,citecolor=blue]{hyperref}
\usepackage[normalem]{ulem}

\usepackage{amssymb,amsmath,amsfonts,amsbsy,graphicx}
\usepackage{color}
\usepackage{psfrag}
\usepackage{cite}

\newcommand\be{\begin{equation}}
\newcommand\ee{\end{equation}}
\newcommand\ba{\begin{eqnarray}}
\newcommand\ea{\end{eqnarray}}\newcommand\eq{\begin{equation}}           
\newcommand\en{\end{equation}}

\input epsf

\usepackage{ulem}
\usepackage{amssymb}
\usepackage{amsmath}
\usepackage{bm}
\usepackage{graphicx}
\usepackage{color}
\usepackage{subfigure}
\usepackage{caption}

\def\gsim{\;\rlap{\lower 2.5pt
 \hbox{$\sim$}}\raise 1.5pt\hbox{$>$}\;}
\def\lsim{\;\rlap{\lower 2.5pt
 \hbox{$\sim$}}\raise 1.5pt\hbox{$<$}\;}

\usepackage{graphicx}
\usepackage{url}
\usepackage{amssymb,amsmath}
\usepackage{amsbsy}

\usepackage{graphicx}
\usepackage{url}
\usepackage{amssymb,amsmath}
\usepackage{amsbsy}

\begin{document}

\begin{titlepage}

\begin{flushright}
CTPU-16-04
\end{flushright}

\begin{center}

{\Large \bf 
A new constraint on millicharged dark matter from galaxy clusters
}\\

\vskip .45in

{
Kenji Kadota,$^a$
Toyokazu Sekiguchi$^a$ and
Hiroyuki Tashiro$^b$
}

\vskip .45in

{\em
$^a$ Institute for Basic Science, Center for Theoretical Physics of the Universe, Daejeon 34051, South Korea \vspace{0.2cm}\\
$^b$ Department of Physics, Graduate School of Science, Nagoya University, Aichi 464-8602, Japan
}

\end{center}

\vskip .4in

\begin{abstract}
We propose a new constraint on millicharged dark matter from considerations 
on galaxy clusters. 
The charged dark matter moves under the influence of the randomly oriented magnetic fields in galaxy clusters, and the corresponding dark matter density profile can significantly differ from the concordance CDM predictions which are well supported from the galaxy cluster observations.
With a typical amplitude of magnetic fields $B=\mathcal O(1)\,\mu$G
and dark matter velocity $v=\mathcal O(100)\,$km/sec at a cluster radius $R\simeq 1$\,Mpc,
we claim that the charge $\epsilon e$ ($e$ is the elementary charge) of dark matter with mass $m$ should 
be bounded as $\epsilon \lesssim 10^{-14}(m/{\rm GeV})$ which is substantially
tighter than the other previous constraints.

\end{abstract}
\end{titlepage}

\setcounter{page}{1}

From the cosmic microwave background to the rotation curves of dwarf galaxies,
the existence of dark matter (DM) is evident in numerous observations at various scales.
Nevertheless, little is known about the nature of DM.
In order to identify DM in the context of particle physics, 
it is crucial to detect its non-gravitational interactions. 
Although DM is widely assumed to be electromagnetically neutral 
in accordance with the cold DM (CDM) paradigm, it is possible
to introduce a particle with a small electric charge $\epsilon e$~(
$\epsilon \ll 1$)~in the Standard Model without breaking the gauge invariance or introducing the anomalies \cite{DeRujula:1989fe}. The quantization of electric charge remains an open question despite the attempts to justify it such as the the search for the magnetic monopoles and the grand unified theories \cite{lang,okun,perl1997}. Such DM with a charge $\epsilon$ much smaller than unity is referred to as the millicharged DM (MDM) and has been searched and tested by the cosmological observations and direct laboratory experiments \cite{sacha,dubo,perl2009,cms,atlas,haas,cora,tasi1,Holdom:1985ag,DelNobile:2015bqo}.
 Such a small charge can make the DM stable due to the charge conservation and also can open up a new avenue for the DM search due to its electromagnetic interactions. 

In particular, for the DM mass $m\gtrsim1\,$GeV, the current tightest constraint on the DM coupling can come from the direct detection experiments (e.g. $\epsilon\lesssim 10^{-11} (m/{\rm GeV})^{1/2}$ from LUX~\cite{Akerib:2013tjd}).

On the other hand, Ref.~\cite{Chuzhoy:2008zy} has pointed out that
the constraints from the direct search experiments should be circumvented by 
taking into account magnetic fields in the Milky Way.
Their argument is twofold. First, MDM 
should have been blown away from the galactic disk by past supernovae in the Milky Way. 
Second, MDM falling from the galactic halo at present should 
undergo cyclotron motion caused by galactic magnetic fields of $\mathcal
O(1)\,\mu$G~\cite{beck,Beck:2013bxa} 
which are largely parallel to the galactic disk. 
The gyroradius of the MDM motion is given by
\begin{equation}
R_g=10^{-9}
\left(\frac{m}{1\,{\rm GeV}}\right) \epsilon^{-1}
\left(\frac{v}{300\,{\rm km\,s}^{-1}}\right)
\left(\frac{B}{1\,\mu {\rm G}}\right)^{-1}\,{\rm pc}, \label{eq:rg}
\end{equation}
where $v$ and $B$ are velocity of DM 
and amplitude of magnetic fields, respectively.
If $R_g$ is smaller than the height of galactic disk ($\simeq100\,$\,pc),
then MDM cannot reach the Earth and direct detection should be ineffective.
These considerations on the galactic magnetic fields can hence lead to the viable MDM parameters: 
$10^{-5} \left({m}/{\rm GeV}\right)\lesssim\epsilon
\lesssim 3\left({m}/{\rm GeV}\right)^{1/2}$ for which the DM direct search bounds are not applicable.

Inspired by the argument above, we in this letter consider the implications of magnetic fields 
in galaxy clusters, whose typical size is $\mathcal O(1)\,$Mpc. 
Although somewhat weaker than galactic ones, magnetic fields have been also observed in many 
galaxy clusters~\cite{Kim:1990}. 
Recent measurements of the Faraday rotation show that the typical magnitude of magnetic fields is 
around $1\,\mu$G~\cite{Govoni:2010yg}, with the radial profile consistent with the hydrodynamic simulations~\cite{Dolag:2000bs}. 
In analogy to MDM in the galactic magnetic fields, MDM at a galaxy cluster scale
should  undergo the cyclotron motion with the gyroradius given in Eq.~\eqref{eq:rg}, on top of drifting via the gravitational force.
We however note that since the orientations of magnetic fields in galaxy clusters are random, the cyclotron motion of MDM
would be more like the random walk with a mean free path being a typical gyroradius.
\footnote{We simply assume the cluster scale magnetic fields are randomly oriented in a cluster. 
This is a reasonable assumption well justified by the simulations and observed polarization properties 
of extended radio sources located inside/behind a galaxy cluster through their Faraday rotation maps~\cite{cari,mur,bona}. 
Our bounds on $\epsilon$ for the galaxy cluster magnetic fields however do not heavily depended on the exact orientations 
of the magnetic fields as long as the magnetic field interactions cannot reproduce the observed cluster scale DM distributions.}
Then it would be reasonable to infer that such DM motion under the influence of the 
magnetic fields can affect the DM distribution in galaxy clusters. The density profile in cluster-sized haloes 
has long been studied both theoretically and observationally. 
Based on the collisionless N-body simulations in the CDM model, 
it has been established that radial distribution of DM in clusters 
to good precision obeys a steep profile. In addition, state-of-art simulations 
incorporating various baryonic processes (e.g. Ref.~\cite{Schaller:2014gwa})
consistently suggest that the distribution of DM can be 
to a good extent fitted by the NFW profile~\cite{Navarro:1995iw} except for very small 
radii ($R\lesssim 10\,$kpc).
These predictions agree well with many observations that directly or indirectly measure 
the cluster density profile at large radii ($R=\mathcal O(100)\,$kpc).
Among them, the gravitational lensing offers, to present, one of the cleanest measurements
of galaxy cluster density profiles. The recent joint analysis of strong 
and weak lensing measurements of stacked clusters~\cite{Umetsu:2015baa}
agrees with CDM predictions to a high degree. The analogous considerations at the galaxy cluster scale apply to
the scenarios with the warm DM (WDM) (typical mass of order
keV) too. The cluster scales ($\gtrsim 1$~Mpc) are larger than the WDM
free streaming scales, and the structure formation proceeds analogously
to those in the CDM scenarios \cite{lov}. The observed DM mass
distribution at the cluster scale of order ${\cal O}(1)~\rm Mpc$ can hence be well described by the conventional predictions assuming the gravitational interactions without the electromagnetic interactions. Having seen that magnetic fields of $B\simeq 1\,\mu$G typically exist in clusters around a radius $R\simeq 1\,$Mpc, we derive the upper bounds on the charge of MDM by looking into the effects of the magnetic fields to be compared with the gravitational effects.

Provided that typical velocity dispersion in clusters is of order $10^3\,{\rm km}\,{\rm s}^{-1}$~\cite{Struble:1999}, which is consistent 
with the NFW profile~\cite{Navarro:1995iw}, we estimate a typical gyroradius, from Eq.~\eqref{eq:rg}, of order $R_g=3\times10^{-15}\epsilon^{-1}(m/{\rm GeV})\,{\rm Mpc}$.
Requiring $R_g\gtrsim1\,$Mpc so that random walk of 
DM should not smear out the density profile within the cluster scale,
we obtain a constraint
\begin{equation}
\epsilon \lesssim10^{-14} \left(\frac{m}{\rm GeV}\right). \label{eq:constraint}
\end{equation}

The analogous bound can also be derived by requiring that the Lorentz force is subdominant compared with the gravitational force.
Let us consider the ratio of the magnitude of the Lorentz force acting on a single DM particle
to the gravitational one, ${\epsilon eBv}/({GM(R)m/R^2})$, 
$M(R)$ is the cluster mass inside a radius $R$ and $G$ is the Newton constant.
Since the centrifugal and gravitational forces balance ($v^2/R \sim GM(R)/R^2$), 
we have $v=10^3\,{\rm km}\,{\rm s}^{-1}$ 
for typical clusters with mass $M(R)=\mathcal O(10^{14})\,M_\odot$ at $R\simeq1\,$Mpc.
Then Eq.~\eqref{eq:constraint} is reproduced from a requirement that 
the Lorentz force is subdominant, ${\epsilon eBv}/({GM(R)m/R^2})<1$.

\begin{figure}
\begin{center}    
\epsfxsize = 0.48\textwidth
\includegraphics{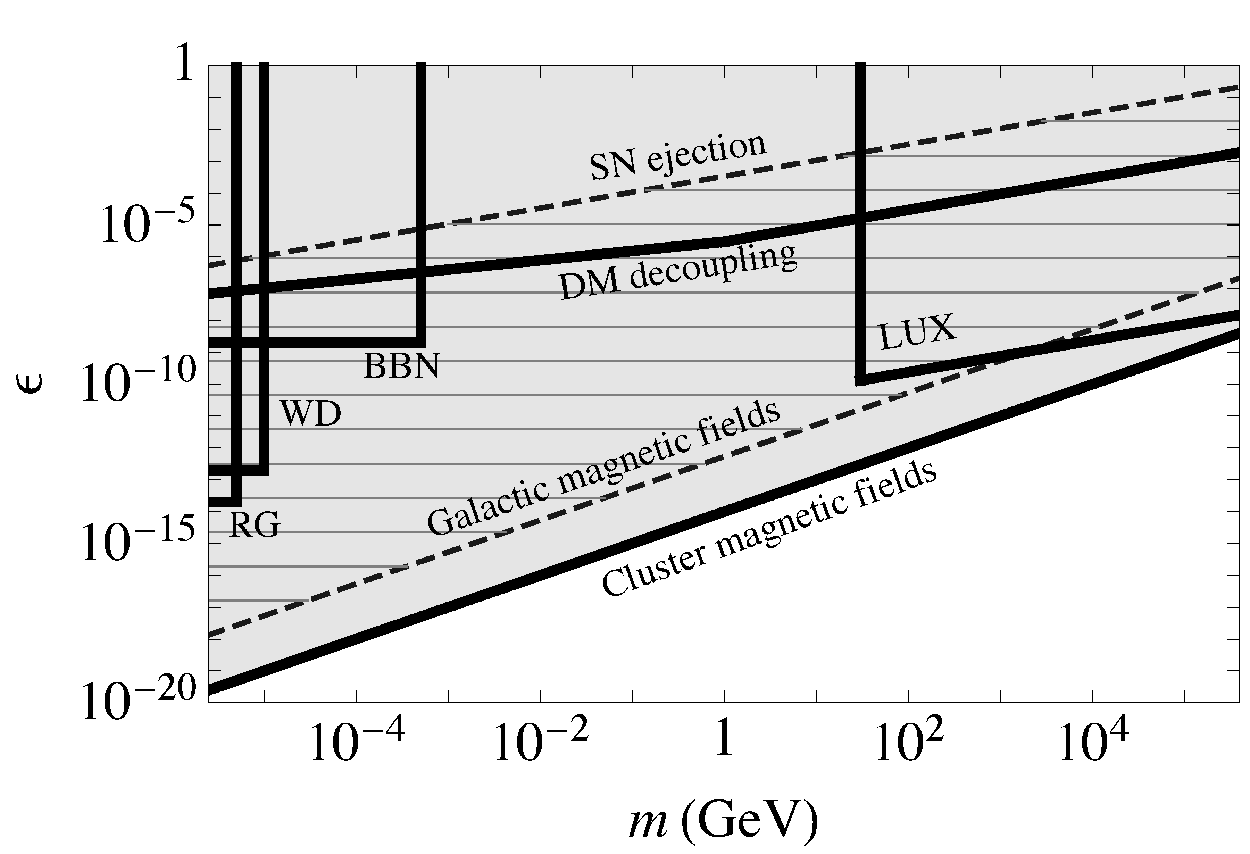}
\end{center}  
\caption{
Constraints on the charge of MDM as a function of its mass. The charge
exceeding the
observational bounds is excluded: Galaxy cluster magnetic fields
affect the cluster scale DM mass distributions and the gray colored
region is hence excluded, which covers the excluded regions considered
previously. The freely streaming
DM over-cools Red Giants and White Dwarfs. Such free streaming of DM is
not applicable for a large $\epsilon$ because of the DM diffusion
while such a large $\epsilon$ is excluded by BBN. The CMB is affected
if the DM does not decouple by the recombination epoch. The direct
DM search bound by LUX is not applicable in the hatched region where
the DM cannot reside in the galactic disk due to the
supernovae and galactic magnetic fields.
 }
\label{jan28milli}
\end{figure}

Our galaxy cluster bound on MDM is shown in Fig. 1. For the comparison, the figure also shows the most stringent bounds from the other observables summarized as follows.
MDM can be produced from the plasmon decay in stars where a photon can acquire an effective mass in a plasma (plasmon). The freely streaming DM can take away the energy and result in too rapid a cooling rate unless $\epsilon$ is small enough to suppress the plasmon decay rate. Constraints from the cooling of the red giant (RG) core and white dwarfs (WD) \cite{sacha} are shown in the figure. Note the bounds from the cooling do not apply for a large DM mass exceeding the plasmon mass for the kinematic reason and also for a large charge ($\epsilon \gtrsim 10^{-8}$ for RG 
 and $\epsilon \gtrsim 10^{-6}$ for WD) due to the trapping of the DM inside the star rather than freely streaming from stars. Those parameter ranges corresponding to the diffusion of DM inside the RG and WD due to a large $\epsilon$ are however excluded from the other stringent constraints such as those from Big-Bang nucleosynthesis (BBN) \cite{sacha}, which is also shown in the figure. MDM can scatter off the nuclei target in the direct DM search experiments, and the current tightest bound coming from the LUX experiment is shown for which the sensitivity to the light DM rapidly decreases for $m \lesssim 30$ GeV due to the small recoil energy~\cite{Akerib:2013tjd,DelNobile:2015bqo}. The bounds from the direct search experiments are not applicable for the parameter range (hatched region) where the millicharged particles are expelled by the supernova explosion shock waves and shielded by the galactic magnetic fields preventing MDM from entering back to the galactic disk \cite{Chuzhoy:2008zy,zurekdark}. The CMB can be affected by the MDM coupling to the photon-baryon fluid because the MDM can influence the CMB analogous to the effects by baryons, and the DM fluctuation growth is also suppressed due to the dragging from the baryon-photon fluid \cite{zurekdark,dubo,bura,cora,tasi1,ali}. There have been hence the stringent constraints on the MDM by demanding that the MDM should decouple from the photon-baryon fluid by the recombination epoch as shown in the figure. The bounds from the other observables such as those from the colliders are weaker or at best comparable to those we just summarized here, for which we refer the readers to the previous literature \cite{sacha,dubo,zurekdark,viny,yavin}.
 One can see that our constraint from the galaxy cluster magnetic fields can substantially improve the previous bounds on the MDM.

Before concluding our discussions, it would be worth mentioning that our constraint does not depend on the coherence scale of magnetic
fields in a galaxy cluster. Even if the coherence scale is smaller than the gyroradius,
the motion of charged DM is easily randomized due to the frequent
change of the magnetic field directions which are randomly oriented at a cluster scale.

 In this letter, we found the simple considerations on the cluster scale magnetic fields affecting the DM distributions in a cluster could lead to the tight bounds on MDM. We derived the quantitative bounds by requiring that the random motion caused by the randomly oriented magnetic fields should not smear out the DM distribution governed by the gravitational interactions and also by demanding that the Lorentz force should not exceed the gravitational force in a cluster. The analytical estimation for the MDM distribution in the halo in existence of the magnetic fields is beyond the scope of this paper analogous to the situation where it is non-trivial and under active debate to analytically derive the NFW-like profile which was obtained by the fitting to the simulation results \cite{dal,wil}. While we postpone the detailed numerical analysis for the MDM distribution in the clusters to our futgure work \cite{prep}, we anticipate that our bounds presented here would be conservative in that the Lorentz force would not necessarily need to exceed the gravitational force to result in the appreciable change in the DM distribution.

\bigskip
\bigskip

\noindent 
\section*{Acknowledgments}

We would like to thank Kiyotomo Ichiki, Kenji Hasegawa and Joe Silk for helpful discussions. K.K. and T.S. are supported by IBS under the project code, IBS-R018-D1.
H.T. is supported by JSPS KAKENHI Grant No.~15K17646 and MEXT's Program
for Leading Graduate Schools PhD professional,``Gateway to Success in Frontier Asia''.


\end{document}